\documentstyle[12pt]{article}

\newcommand{\bbr}{I\!\! R}
\newcommand{\bbz}{Z\!\!\! Z}

\begin{document}
\thispagestyle{empty}
\begin{center}

\null
\vskip-1truecm
\vskip2truecm
{\bf STRINGY INSTABILITY OF TOPOLOGICALLY NON-TRIVIAL AdS BLACK HOLES AND OF deSITTER S-BRANE SPACETIMES\\}
\vskip2truecm
Brett McInnes
\vskip2truecm

Department of Mathematics, National University of Singapore, 10 Kent Ridge Crescent,
Singapore 119260, Republic of Singapore.\\ 
E-mail: matmcinn@nus.edu.sg\\    

\end{center}
\vskip1truecm
\centerline{ABSTRACT}
\baselineskip=15pt
\medskip

Seiberg and Witten have discussed a specifically ``stringy" kind of instability which arises in connection with ``large" branes in asymptotically AdS spacetimes. It is easy to see that this instability actually arises in most five-dimensional asymptotically AdS black hole string spacetimes with non-trivial horizon topologies. We point out that this is a more serious problem than it may at first seem, for it cannot be resolved even by taking into account the effect of the branes on the geometry of spacetime. [It is ultimately due to the {\em topology} of spacetime, not its geometry.]  Next, assuming the validity of some kind of dS/CFT correspondence, we argue that asymptotically deSitter versions of the Hull-Strominger-Gutperle S-brane spacetimes are also unstable in this ``topological" sense, at least in the case where the R-symmetries are preserved. We conjecture that this is due to the unrestrained creation of ``late" branes, the spacelike analogue of large branes, at very late cosmological times. 
\vskip3.5truecm
\begin{center}

\end{center}

\newpage

\addtocounter{section}{1}
\section*{1. Large Brane Instability and Topologically Non-Trivial AdS Black Holes}
The fact that asymptotically AdS black holes can have topologically non-trivial event horizons \cite{kn:lemos}\cite{kn:peldan}\cite{kn:mann} has attracted considerable attention in connection with the AdS/CFT correspondence. In five dimensions \cite{kn:birmingham}, apart from the spherical, flat, and hyperbolic possibilities for the structure of the event horizon, Milnor's \cite{kn:milnor} ``prime decomposition" of compact orientable 3-manifolds suggests further candidates, and black hole spacetimes with some of the corresponding event horizons have been constructed in a remarkable paper of Cadeau and Woolgar \cite{kn:cadeau}. 

In \cite{kn:witten2}, Witten remarks that if one wishes to generalise the AdS/CFT correspondence to bulk manifolds which are merely {\em asymptotically} AdS, then [in the Euclidean formulation at least] one must take care that the scalar curvature at conformal infinity should remain non-negative. For otherwise the massless scalars of the CFT will undergo ``runaway" behaviour. For AdS space itself, the [Euclidean] boundary is a conformal sphere with its canonical structure, represented by a metric of positive scalar curvature; the latter prevents a runaway through the conformal coupling term. Notice that it is {\em not} claimed that {\em all} CFTs misbehave on spaces of negative curvature -- this is of course not the case. The instability arises only for a subclass of CFTs arising ``holographically", as for example in the AdS/CFT correspondence. [This point is emphasised in \cite{kn:witten3}; see also \cite{kn:yau}]. Note that the two-dimensional case is particularly delicate here; but we stress that, in any case, {\em none of the CFTs discussed in this work will be defined on a manifold of dimension less than three}.

The bulk version of this runaway was explained in detail in \cite{kn:seiberg}; the instability arises from the nucleation and growth of ``large branes", that is, in the five-dimensional case, 3-branes which are generated at large distances, ``near" the boundary. The relevant Euclidean action for a (D-1)-brane in a (D+1)-dimensional asymptotically AdS space is, in the notation of \cite{kn:seiberg},
\begin{equation} \label{eq:action}
S =  {Tr^D_0 \over 2^D} \int \sqrt{g}\; \left( (1-q)\phi^{{2D\over D-2}}+{8 \over (D-2)^2}[(\partial\phi)^2 + {D-2\over 4(D-1)}\phi^2R] + {\cal O}(\phi^{{2(D-4)\over D-2}})\right). 
\end{equation}
Here $D$ is {\em strictly} greater than 2 --- that is, this equation, and consequently the rest of our discussion, is only valid for spacetimes of dimension at least 4 [and consequently only for conformal boundaries of dimension at least 3]. The brane carries a charge $q$ under a background antisymmetric field; the field ${\phi}$ tends to infinity as the conformal boundary is approached, and $R$ is the scalar curvature of that boundary. In the BPS case, the action will be unbounded below if $R$ is negative: there will be unstable production of branes ``near" the boundary. The case where $R$ is zero is more delicate; one expects instability in some cases but not in others, depending on higher-order corrections \cite{kn:witten3}. As the boundary metric is only defined modulo conformal factors, and as scalar curvature is not a conformal invariant, perhaps we should clarify. A conformally invariant formulation is most easily given in terms of the scalar fields on the boundary. These are governed by the ``conformal Laplacian", defined [for $D > 2$] by
\begin{equation} \label{eq:conflap}
\Delta_{CON} = \Delta + {{D-2} \over {4(D - 1)}}R,
\end{equation}
where $\Delta$ is the ordinary Laplacian. The conformally invariant form of the stability condition is that this operator should be non-negative \cite{kn:yau}. For this to hold, it must be possible to choose a conformal gauge such that $R$ [a function of position in general] is everywhere non-negative. In the the case of a {\em compact} boundary, this is easier to understand because it is then always possible \cite{kn:parker} to choose a conformal gauge such that $R$ is {\em constant}. This is usually, but not invariably \cite{kn:jin}, also possible for complete, non-compact Riemannian manifolds.

Now consider a five-dimensional Euclidean AdS black hole with metric
\begin{equation} \label{eq:adshyper}
g(EAdS_5;hyper) = {f_5}(r) \; dt \otimes dt + {f_5}^{-1}(r) \; dr \otimes dr + r^2h_{ij}[H^3/\Gamma] \; dx^i \otimes dx^j
\end{equation}
with
\begin{equation} \label{eq:f}
{f_5} = -1 - {16\pi G \mu \over 3r^2 \; {\rm{Vol}}({H^3}/ {\Gamma})} + {r^2 \over L^2},
\end{equation}
where ${\mu}$ is the mass parameter and $h_{ij}[H^3/\Gamma] \; dx^i \otimes dx^j
$ is the metric on the 
three-dimensional horizon, ${H^3}/ {\Gamma}$. Here $H^3$ is the three-dimensional hyperbolic space of constant sectional curvature $-1$, and $\Gamma$ is an infinite discrete group acting on $H^3$ in such a way that the quotient is compact. The metric at infinity [that is, the canonical representative of the induced conformal structure on the boundary] is
\begin{equation} \label{eq:gamma4}
{\gamma}_4 = dt \otimes dt + {L^2}h_{ij}[H^3/\Gamma] \; dx^i \otimes dx^j.
\end{equation}
The scalar curvature of this metric is $-6/{L^2}$. Thus, large branes apparently grow uncontrollably at large distances in this spacetime. Before commenting on this, let us take note of the topology of the space on which ${\gamma}_4$ is defined. First, as usual for Euclidean black holes, we have to identify time periodically, so $t$ is a parameter on a circle. Next, the topology of hyperbolic space is that of ${\bbr}^3$, so we see finally that the topology of the boundary is that of ${S^1} \times {{\bbr}^3}/{\Gamma}$. Of course, the conformal boundary of Euclidean AdS space itself has a very different topology, namely that of a sphere. Thus we see that a spacetime can be asymptotically AdS, {\em and yet have a conformal infinity with a topology which differs very radically from that of the boundary of AdS space}.

It is clear that the hyperbolic AdS black hole is not stable as a string background. However, we wish to argue that it is by no means obvious, {\em from this alone,} that such spacetimes must be rejected as string backgrounds. Instability, in reality, rarely leads to indefinitely increasing perturbations: it is usually ultimately self-limiting, and we should assume that this is the case here unless we can prove the contrary. In using the Seiberg-Witten criterion in the above way, we are ignoring the effect of the branes on the space-time geometry. This is a good approximation at short distances, but eventually, at extremely large distances from the event horizon, the branes must have such an effect. In the Lorentzian picture, the geometry of the spacelike sections will begin to evolve in response to the branes, and, as can be seen from the form of equations 3 and 5 [in which the geometry of the spacelike sections is clearly influenced   by the geometry at infinity and vice versa], this means that the geometry at infinity must also change. This will in general change the scalar curvature, so the Seiberg-Witten instability condition might cease to hold. In physical language, one would say that the back-reaction of the branes on the geometry eventually causes the spacetime to settle down to a new, approximately static state. But now we have an obvious consistency check: clearly, the scalar curvature of the boundary of the [Euclidean version of] the final spacetime geometry must be positive or zero; otherwise we would have a contradiction. In the Euclidean language of \cite{kn:seiberg}: in order for large-brane instability to limit itself through back-reaction, it must be possible to deform the Euclidean version of the spacetime in such a way that the scalar curvature at infinity changes from negative to non-negative values. In short, {\em unless we can prove otherwise,} we should assume that the real significance of Seiberg-Witten instability for this spacetime is that formulae \ref{eq:adshyper} and \ref{eq:f} are not to be trusted at extremely large distances, so that the metric $\gamma_4$ does not accurately reflect the geometry of the boundary in this case. This greatly complicates the study of these black holes from the AdS/CFT point of view, but in itself it does not rule them out as string backgrounds.

Circumstantial evidence that something of this kind can happen comes from the study of asymptotically AdS black holes with {\em flat}, toral event horizons. Intuitively, one would expect that AdS itself should be the thermal background for any asymptotically AdS black hole, and this is certainly the case when the horizon is spherical. But it has been argued \cite{kn:surya} that it is {\em not} so when the horizon is a torus. The argument is based on a conjecture, due to Horowitz and Myers \cite{kn:horowitz}, that the ``AdS soliton" is the lowest energy metric among all metrics which are asymptotic to it sufficiently rapidly. The [Euclidean] metric of the soliton is given, in $n+1$ dimensions, by
\begin{equation} \label{eq:adssoliton}
g(AdS;soliton) = {r^2}dt \otimes dt + {{L^2} \over {r^2}}{(1 - {{{r_0}^n} \over {r^n}})^{-1}}dr \otimes dr + {{r^2} \over {L^2}}{(1 - {{{r_0}^n} \over {r^n}})}d\phi \otimes d\phi + {r^2}\sum^{n-2}_{i=1} \; d\theta_i \otimes d\theta_i. 
\end{equation}
Here $r_0$ and L are positive constants, $r$ is a radial coordinate satisfying $r>r_0$, and $\phi$ and the $\theta_i$ are angular coordinates of various periodicities. In the Euclidean case, $t$ too is periodic. Thus the conformal infinity of the soliton has the structure of an n-dimensional torus, $T^n$. This is of course the same as the conformal infinity of an asymptotically AdS black hole with a flat [toral] horizon, and the black hole metric does approach that of the soliton sufficiently rapidly for the Myers-Horowitz conjecture to apply. If the latter is correct, then, as argued in \cite{kn:surya}, we should certainly use the soliton as the thermal background for toral asymptotically AdS black holes, {\em not} AdS itself.

There is now quite impressive evidence in favour of the Myers-Horowitz conjecture \cite{kn:galloway1}\cite{kn:galloway2}\cite{kn:gallowaynew} and for the claim that the soliton is the correct thermal background for toral black holes \cite{kn:page1}\cite{kn:page2}. The resulting picture of these ``toral-boundary" spacetimes is very satisfactory; for example, they exhibit a well-defined confinement/deconfinement transition similar to, but interestingly different from, the more familiar transition which is known to occur \cite{kn:witten4} in the case of the AdS-Schwarzschild black hole. There is certainly no hint of any kind of runaway on the boundary or large-brane instability in the bulk. Now the conformal structure at infinity is represented by an exactly flat metric. But suppose that we slightly perturb the geometry of either the toral black hole or the soliton at some point deep in the bulk. Suppose that this perturbation causes the boundary conformal structure to change so that it is no longer the conformal structure represented by the flat metric. Then, using techniques to be explained below, one can show that the scalar curvature of the resulting metric on the boundary {\em cannot be everywhere positive or zero}. That is, the slightest perturbation of the boundary conformal geometries of these spacetimes renders them unstable in the Seiberg-Witten sense.

We suggest that this can be reconciled with the good thermodynamical behaviour of these spacetimes by invoking back-reaction as above. That is, we predict that the [Lorentzian] geometry evolves in such a way that the [Euclidean version of] the final metric corresponds to a boundary metric with a scalar curvature that has been driven back towards zero. [It cannot become positive everywhere --- see below.] If this prediction proves to be false, then the implication is that the AdS soliton is unstable against arbitrarily small fluctuations, which seems very unlikely. [Note that it follows from the main theorem stated in the next section that {\em every} metric of zero scalar curvature on a torus is perfectly flat --- that is, if the scalar curvature vanishes on a torus, so does every curvature component. This does not imply, however, that the final metric after back-reaction effects is necessarily identical to that given in equation 6. For it is known \cite{kn:anderson} that when the boundary is topologically a torus, the boundary conformal structure does not uniquely determine the bulk metric, not even when the bulk is an Einstein manifold.]
 
The question, then, is this: can the manifold ${S^1} \times {{\bbr}^3}/{\Gamma}$, which we obtained above as the Euclidean boundary of the five-dimensional hyperbolic AdS black hole spacetime, be given a metric of non-negative scalar curvature? Intuition may suggest that it should be impossible to find a metric of positive scalar curvature on ${{\bbr}^3}/{\Gamma}$, but, before we trust intuition, note that it is certainly possible [\cite{kn:besse}, page 123] to find a metric of {\em constant negative} scalar curvature on the 3-sphere $S^3$. The point is that the scalar curvature is just the ``average" of all of the curvature components. Thus, if even one component can be made sufficiently negative at each point, the scalar curvature on $S^3$ will be negative at each point. The reader can visualise this by imagining a deformation of the ordinary 3-sphere such that in {\em some} directions at each point the geometry becomes ``saddle-like", while remaining ``sphere-like" in other directions. [This corresponds to modifying the metric, but {\em not} the topology, of the 3-sphere.] Similarly, while it is of course impossible to force {\em all} of the sectional curvatures on ${{\bbr}^3}/{\Gamma}$ to be positive, one would be surprised to find that this cannot be arranged for the {\em average}. Even if it is indeed impossible to construct a positive scalar curvature metric on ${{\bbr}^3}/{\Gamma}$, it might still be possible to construct a warped product [or even more general] metric on ${S^1} \times {{\bbr}^3}/{\Gamma}$ with positive scalar curvature, since warping can certainly change the sign of scalar curvature $-$ compare the positive scalar curvature product metric $dx \otimes dx + d\theta \otimes d\theta + {sin^2}(\theta)d\phi \otimes d\phi$ on $(0, \infty) \times S^2$ with the negative scalar curvature warped metric $dx \otimes dx + {sinh^2}(x) [d\theta \otimes d\theta + {sin^2}(\theta)d\phi \otimes d\phi]$. Clearly we need some powerful mathematical technique to settle this. 

Such a technique exists, and in the next section we give a very brief overview of it. The upshot is that it is actually {\em impossible} to define a metric of positive or zero scalar curvature on this manifold: no matter how we deform it, the scalar curvature remains resolutely negative. Thus, these black hole spacetimes are unstable [if we embed them in string theory] in a very radical way. The only escape route is to suppose that the branes actually change the {\em topology} of spacetime, a possibility that we shall also consider. Finally, in section 3 we
show that, if some kind of ``dS/CFT" correspondence is valid, then any asymptotically deSitter
version of the Hull-Strominger-Gutperle ``S-brane" spacetime is also radically unstable in the same sense, at least if we maintain the ansatz --- used to obtain all known examples of S-brane spacetimes --- that the R-symmetries are preserved. We believe that this is due to ``late" S-branes, the spacelike analogue of ``large" branes.

\section*{2. Topologically Induced Instability}

In this section we introduce a geometric technique which, in view of the importance of scalar curvature in Seiberg-Witten instability, is the natural one for our purposes. The main reference is \cite{kn:lawson}.

A smooth map $f: X \rightarrow Y$ from one Riemannian manifold to another is said to be $\epsilon$-contracting if for any piecewise smooth curve $C$ in $X$, 
\begin{equation} \label{eq:curve}
L[f(C)] \leq {\epsilon} \times L[C],
\end{equation}
where $L$ denotes the length. A map which takes every point of $X$ to a single point of $Y$ evidently satisfies this condition for any positive $\epsilon$. To avoid this trivial case, we consider only mappings of non-zero {\em degree} [see \cite{kn:lawson}, page 303; note that mappings may have to be ``constant at infinity" to make sense of this if $X$ happens to be non-compact.] An n-dimensional manifold $M$ is said to be {\em enlargeable} if, given any Riemannian metric on $M$, for any positive $\epsilon$ there exists an orientable covering manifold which admits an $\epsilon$-contracting map [with respect to the pull-back metric] of non-zero degree onto the unit n-sphere. Notice that enlargeability is a topological condition; we can nevertheless think of an enlargeable manifold as one which, like a torus but unlike a sphere, has ``arbitrarily large" covering spaces. Simple examples are provided by compact manifolds, such as tori and the underlying manifolds of compact hyperbolic spaces, which admit a metric of non-positive {\em sectional} curvature. Thus, ${{\bbr}^3}/{\Gamma}$ above is enlargeable. The following properties of enlargeable manifolds are important. [See \cite{kn:lawson}, page 306.]

\begin{enumerate}
\item The product of any compact enlargeable manifold with a torus of any dimension is again          enlargeable.
\item The connected sum of a compact enlargeable manifold with any compact manifold is again enlargeable.
\item If the scalar curvature of a Riemannian metric on a compact enlargeable spin manifold is non-negative, then the metric must be flat.
\end{enumerate}

[Recall that the connected sum of two manifolds is obtained by removing small balls from each, and then joining the two along the resulting boundaries.] Now point 1 informs us that ${S^1} \times {{\bbr}^3}/{\Gamma}$ is enlargeable. Point 3 then implies, since this space [like all products of a circle with an orientable compact 3-dimensional manifold] is a spin manifold, that there can be no metric of positive scalar curvature on this manifold, {\em not even if we allow warped products.} Furthermore, if the scalar curvature were zero, then the metric would have to be flat. But this is not possible: ${S^1} \times {{\bbr}^3}/{\Gamma}$ does not have the topology of a flat manifold. [One says that $\Gamma$ is {\em homotopically atoroidal} [\cite{kn:besse},page 158]  and that, as the name suggests, means that $\Gamma$ cannot occur as part of the fundamental group of a manifold which can be flat \cite{kn:wolf}.] Thus, the scalar curvature of ${S^1} \times {{\bbr}^3}/{\Gamma}$ {\em cannot be everywhere positive or zero}, no matter how the manifold is deformed.

We conclude that once large branes begin to develop in the hyperbolic AdS black hole spacetime, nothing can rein in the instability: for, no matter how the branes deform the spacetime, the scalar curvature at infinity can never become everywhere positive or zero. The instability is {\em induced topologically} \cite{kn:mcinnes2}. We are forced to conclude that these spacetimes simply cannot arise as solutions of string or M theory. In fact, this conclusion is actually consistent with the findings of \cite{kn:surya}, where the authors remark that {\em no analogue of the AdS soliton seems to exist in the hyperbolic case}. It is now clear why this is so: there can be no well-behaved ground state, analogous to the AdS soliton, for these radically unstable spacetimes. [We agree with these authors that pure AdS is not the correct ground state to use in the hyperbolic case; this is argued most convincingly in the introduction to \cite{kn:galloway2}.] 

More generally, Milnor's \cite{kn:milnor}  ``prime decomposition" theorem states that any compact orientable 3-manifold $M^3$ can be expressed as a connected sum in the following way:
\begin{equation} \label{eq:hex}
M^3 = {\Sigma_1} \# {\Sigma_2} \# ... \# (S^1 \times S^2) \# (S^1 \times S^2) \# ... \# K_1 \# K_2 \# ...,
\end{equation}
where each $\Sigma_i$ is a manifold covered by a homotopy 3-sphere, where $\#$ denotes the connected sum, and where each $K_i$ is an Eilenberg-MacLane space of the form $K(\pi,1)$. We shall assume the truth of the Poincar\`e conjecture, so that the reader can interpret ``homotopy 3-sphere" as $S^3$. Then the $\Sigma_i$ are just quotients of $S^3$ by [completely known] finite groups; for example, the real projective space $S^3/{\bbz_2}$ is one possibility. [See \cite{kn:weeks} for a readily accessible statement of the classification.] A $K(\pi,1)$ space is just a 3-dimensional manifold whose only non-trivial homotopy group is its fundamental group. It is not yet proven that all compact $K(\pi,1)$ spaces are enlargeable, but all known examples are so, and furthermore it is known [\cite{kn:lawson}, page 324] that no such manifold can accept a metric of positive scalar curvature, and that the same is true of the connected sum of a $K(\pi,1)$ with any other compact manifold. Therefore, in view of properties 2 and 3 listed above, it is reasonable to conjecture that all compact $K(\pi, 1)$ spaces are enlargeable.

Assuming the truth of this, we see that any compact orientable 3-manifold is enlargeable if it has at least one $K_i$ in its Milnor decomposition. Thus we find that compact orientable 3-manifolds fall into three categories: the six well-understood \cite{kn:wolf} manifolds which can be flat [see \cite{kn:reboucas} for an accessible review], the enlargeable manifolds of ``non-flat" topology, and manifolds of the form ${\Sigma_1} \# {\Sigma_2} \# ... \# (S^1 \times S^2) \# (S^1 \times S^2) \# ...$ Every member of this last class has a metric of constant positive scalar curvature \cite{kn:schoen}\cite{kn:joyce}. Assume that, as is the case in physically interesting examples, the topology of the black hole spacetime is a product of the event horizon topology with ${\bbr}^2$. Then the conformal boundary of the corresponding Euclidean asymptotically AdS black hole spacetime in five dimensions will be a product of $S^1$ with a member of one of these classes, and so property 1 above implies that the scalar curvature of the conformal boundary cannot be non-negative unless the event horizon is either flat or of the form ${\Sigma_1} \# {\Sigma_2} \# ... \# (S^1 \times S^2) \# (S^1 \times S^2) \# ...$. Hence we conclude that large-brane instability rules out all five-dimensional asymptotically AdS black holes [as string backgrounds] except those with flat or ${\Sigma_1} \# {\Sigma_2} \# ... \# (S^1 \times S^2) \# (S^1 \times S^2) \# ...$ event horizons. [Strictly speaking, this is under the assumption that the spacetime topology has the above product form. We conjecture, however, that it is true in general that the conformal boundary of [the Euclideanized version of] a black hole spacetime with an enlargeable event horizon is necessarily itself enlargeable.] The flat case we have considered already, and the AdS-Schwarzschild black hole can have any $\Sigma_i$ as event horizon. It would be interesting to exhibit a black hole --- perhaps one should really say ``black string" --- with, for example, ${\Sigma_1} \# {\Sigma_2}  \# (S^1 \times S^2)$ as event horizon. In any case, these are the only remaining possibilities for event horizons of string theoretic asymptotically AdS black holes in five dimensions.

Throughout this discussion, we have assumed that while the nucleation of large branes can change the geometry of spacetime, it cannot change the {\em topology}. But it is well known that D-branes can in fact change the topology of 10 or 11 dimensional spacetimes in string/M theory \cite{kn:aspinwall1}\cite{kn:aspinwall2}. However, that kind of topology change usually involves changes of topology among the members of some {\em family} of spacetimes as a parameter is changed. Even in the cases where the topology change occurs within {\em one} spacetime, as in \cite{kn:greene}, it occurs along some spacelike dimension of a higher-dimensional spacetime. In our case, in order to affect the topology at infinity, the branes would have to change the topology of the spacelike sections as they evolve in {\em time} in the Lorentzian version of the spacetime. [Again, to see this, consider equations 3 and 5: the topology of the boundary is controlled by that of the spatial sections.] But classical theorems of Geroch \cite{kn:geroch} and Tipler \cite{kn:tipler} state that spacetimes in which topology changes occur due to temporal evolution are singular or contain closed timelike worldlines. Admittedly, Horowitz \cite{kn:horowitz2} has argued that the singularities are not necessarily very drastic in cases where the topology change is very simple. Even if we accept that argument, however, the topology change here would necessarily be very extreme --- from a compact hyperbolic space, for example, with its infinite and very complex fundamental group, to a sphere. It is most unlikely that such a major topological change can be effected without inducing non-innocuous singularities or causality violations. In short, it is conceivable that that large branes can avert a catastrophic runaway by modifying the topology of spacetime, but it is highly likely that this has to be paid for by some kind of equally catastrophic gravitational collapse in the region far {\em outside} the event horizon. Thus it seems clear that changing topology cannot save the situation, though the details of this deserve further investigation.  

We wish to impress on the reader the extreme nature of the restriction imposed on black hole horizons by Seiberg-Witten instability. For there is a definite sense in which compact orientable 3-manifolds having no $K(\pi, 1)$ component in their Milnor decomposition are a ``tiny minority". This instability rules out a large number of apparently acceptable spacetimes. We now argue that, if there is a ``dS/CFT correspondence", it has a similar effect in the case of asymptotically deSitter S-brane spacetimes.

\addtocounter{section}{3}
\section*{3.  Late Brane Instability of Asymptotically deSitter S-Brane Spacetimes}

[Throughout this section, for technical reasons connected with the form of the S-brane metric we shall consider below, all spacetimes will be four-dimensional; therefore all event horizons will be two-dimensional and all CFTs will be defined on {\em three-dimensional} conformal boundaries.]

The above discussion of the Seiberg-Witten instability was given, as in \cite{kn:seiberg}, in the Euclidean formulation. To repeat: a negative scalar curvature on the boundary gives rise to a scalar field ``runaway" which is the AdS/CFT counterpart of the unstable emission of ``large branes" in the bulk. When we turn to the dS/CFT correspondence [\cite{kn:hull1}\cite{kn:park1}\cite{kn:park2}\cite{kn:witten1}\cite{kn:strominger1}\cite{kn:strominger2}\cite{kn:medved1}; see \cite{kn:balasubramanian}\cite{kn:leblond} for relevant recent discussions], we find that, since conformal infinity is now spacelike, the CFT is automatically defined on a space of Euclidean signature, even if the bulk is Lorentzian. We can therefore reasonably hope that the lessons we learned in the Euclidean AdS case will apply here: a CFT on a boundary with scalar curvature which is not positive or zero will suffer a runaway in this case too. [We should perhaps stress that the dS/CFT correspondence is by no means supported by as much evidence as its AdS counterpart. However, for our purposes, the precise details, and even the exact validity, of the dS/CFT correspondence are not essential. All we require is that a runaway in a CFT at infinity should be reflected in some kind of bulk instability. We certainly expect this to hold true even if, as Susskind \cite{kn:susskind} has suggested, the dS/CFT correspondence can only be approximately valid. In fact, anything else would signal a basic failure of the holography principle in the deSitter context.] 

Now in fact {\em non-singular} asymptotically deSitter spacetimes are well-behaved from this  point of view. For the main result of \cite{kn:galloway3} states the following. Suppose that   an asymptotically deSitter spacetime is {\em future asymptotically simple}, in the sense that every future-directed inextendible null geodesic has an end point on future null infinity; this essentially just means that there are no singularities. Then, if the Dominant Energy Condition [DEC] holds, the spacetime is globally hyperbolic with compact Cauchy surfaces having finite fundamental groups. This result implies that future and past conformal infinity are also compact with finite fundamental groups. [Notice that the DEC is essential here: the spacetime given in \cite{kn:mcinnes1} is asymptotically deSitter and future asymptotically simple, but conformal infinity has an infinite fundamental group. Of course, this spacetime does not satisfy the DEC.] In the case of a four-dimensional asymptotically deSitter space, this compactness in turn means that [assuming orientability] we can use the Milnor decomposition; and clearly there can be no $K(\pi, 1)$ factors in this case, since such manifolds always have infinite fundamental groups. [This does {\em not} mean that the scalar curvature must necessarily be positive $-$ recall that the scalar curvature {\em can} be negative even for a 3-sphere $-$ but it does mean that the manifold can be deformed so that the scalar curvature {\em becomes} positive, that is, all these manifolds admit a metric of positive scalar curvature \cite{kn:schoen}\cite{kn:joyce}.] Thus, there is no danger of Seiberg-Witten instability for non-singular asymptotically deSitter spacetimes.

Of course, this result does not apply to spacetimes that are asymptotically deSitter but not future asymptotically simple. To see why such spacetimes could be of considerable interest, consider the {\em spacelike branes} introduced by Hull \cite{kn:hull2} and analysed in detail in \cite{kn:gutperle1} [see also \cite{kn:gutperle2}\cite{kn:peet}\cite{kn:ivashchuk}; see \cite{kn:hashimoto} for more recent references]. An explicit four-dimensional S-brane metric is given by \cite{kn:gutperle1} 
\begin{equation} \label{eq:ansatz}
-{{d\tau^2}\over\lambda^2} + {\lambda^2}dz^2 + R^2 d{H_2}^2,
\end{equation}
where $d{H_2}^2$ denotes a locally hyperbolic metric [of constant curvature $-1$], and where $\lambda$ and $R$ are simple functions of the time coordinate $\tau$. [The hyperbolic term ensures the preservation of the R-symmetries.] This metric is asymptotically {\em flat} in the remote past and future. Its Penrose diagram \cite{kn:buchel1}\cite{kn:buchel2} is obtained simply by means of a ninety degree rotation of the Penrose diagram of a Schwarzschild black hole, with the understanding that each point represents a locally hyperbolic space instead of a sphere. Thus the spacetime has non-spacelike future and past conformal boundaries and is nakedly singular. [The significance of this last property is discussed in \cite{kn:buchel1}\cite{kn:buchel2}. See also \cite{kn:burgess}\cite{kn:quevedo} for an interesting interpretation of the singularities.] Thus we have an important example of a spacetime which is not future asymptotically simple. 

As our Universe is probably \cite{kn:ratra} asymptotically deSitter rather than asymptotically flat, it is very important to ask whether it is possible to modify the S-brane geometry so that the conformal boundary becomes spacelike. If such asymptotically deSitter S-brane spacetimes exist, their Penrose diagrams will be square, with naked singularities to each side as above, but with horizontal [spacelike] upper and lower boundaries. That is, their Penrose diagrams will be obtained from a ninety degree rotation of the diagram of an AdS black hole, just as the Penrose diagram of the above metric was obtained from a ninety degree rotation of the diagram of a Schwarzschild black hole. Let us try to construct an explicit example of a metric with such a Penrose diagram. [The intention here is not to find an S-brane spacetime, but rather to give a concrete example of a metric with such a Penrose diagram, so that the structure of future conformal infinity can be determined for any spacetime having a diagram of this type. Following \cite{kn:buchel1}\cite{kn:buchel2}, we maintain the R-symmetries by assuming that the horizons are hyperbolic.] 

The Lorentzian four-dimensional anti-deSitter black hole with a hyperbolic horizon has metric
\begin{equation} \label{eq:adshyper4}
g(AdS_4;hyper) = - {f_4}(r) \; dt \otimes dt + {f_4}^{-1}(r) \; dr \otimes dr + r^2h_{ij}[H^2/\Delta] \; dx^i \otimes dx^j
\end{equation}
with
\begin{equation} \label{eq:f4}
{f_4} = -1 - {8\pi G \mu \over r \; {\rm{Vol}}({H^2}/ {\Delta})} + {r^2 \over L^2};
\end{equation}
note that the event horizon is now a compact Riemann surface, $H^2/\Delta$, where $\Delta$ is a discrete infinite group acting freely and properly discontinuously on the hyperbolic plane, $H^2$. [ The conformal boundary is then a three-dimensional manifold with topology given by the product of the Riemann surface topology with a line.] The Penrose diagram, as stated above, is a square with horizontal black and white hole singularities and with vertical timelike infinities. Now simply rotate this diagram through 90 degrees. The new diagram has spacelike, horizontal infinities and timelike, naked singularities to either side: it resembles what in the West would be called a Chinese lantern. This is the desired nakedly singular asymptotically deSitter spacetime; the AdS event horizon has been transformed to the dS cosmological horizon. Notice that, in equation  \ref{eq:adshyper4}, the spacelike and timelike roles of $r$ and $t$ are exchanged inside the event horizon as usual. We can use this to deduce the metric of the ``rotated" spacetime as follows. If $\alpha$ denotes the radius of the event horizon, then set $r = \alpha - \epsilon$ and substitute this into equation \ref{eq:f4}. The result is, up to first order in $\epsilon$,
\begin{equation} \label{eq:epsilon}
{f_4} = {8\pi G \mu \epsilon \over {\alpha^2} \; {\rm{Vol}}({H^2}/ {\Delta})} + {{2\alpha\epsilon}\over L^2}.
\end{equation}
Now we want the geometry just {\em inside} the AdS {\em event} horizon to resemble the geometry just {\em outside} the dS {\em cosmological} horizon. If the parameters are adjusted so that the latter has radius $\alpha$, then we have $r = \alpha + \epsilon$ in this case, and so we see that the desired metric is obtained simply by reversing the signs of the last two terms in equation \ref{eq:f4}. That is, the dS analogue of the AdS black hole with hyperbolic event horizon is the nakedly singular spacetime with metric
\begin{equation} \label{eq:dshyper}
g(dS_4;hyper) = - j(r) \; dt \otimes dt + j^{-1}(r) \; dr \otimes dr + r^2h_{ij}[H^2/\Delta] \; dx^i \otimes dx^j
\end{equation}
where
\begin{equation} \label{eq:j}
j = -1 + {8\pi G \mu \over r \; {\rm{Vol}}({H^2}/ {\Delta})} - {r^2 \over L^2}.
\end{equation}

Of course we are not suggesting that an asymptotically deSitter S-brane metric will have exactly this form: no doubt there will be differences. The point, however, is that, in view of the way we constructed it, we can expect the {\em asymptotic} behaviour of this spacetime to mimic that of an asymptotically deSitter S-brane spacetime. Since we propose to use the dS/CFT correspondence, this is all we need. 

Now in fact the metric 13 has already been discussed in the dS/CFT literature: it was introduced in \cite{kn:cai1} as an example of a ``topological deSitter" spacetime. In fact, the hyperbolic deSitter spacetime was found to be particularly well-behaved, in that it has a positive Cardy-Verlinde Casimir energy \cite{kn:cai2}\cite{kn:morecai}\cite{kn:medved3}. Thus, the situation here is particularly favourable from the dS/CFT point of view. Let us therefore investigate the structure of conformal infinity. Outside the cosmological horizon, the coordinate $r$ in equation \ref{eq:dshyper} becomes timelike, and so the geometry of conformal infinity is revealed by letting $r$ tend to infinity. In view of the way we derived the metric, it is no surprise to find that the canonical representative of the conformal structure is almost precisely the same as in the case of the Euclidean hyperbolic AdS black hole, that is, it is given by 
\begin{equation} \label{eq:gamma3}
{\gamma}_3 = dt \otimes dt + {L^2}h_{ij}[H^2/\Delta] \; dx^i \otimes dx^j.
\end{equation}
Here $\gamma_3$ is a metric on a three-dimensional Euclidean manifold [since $t$ has become spacelike] with scalar curvature $-2/L^2$. The full conformal boundary consists of two copies of this manifold, one in the infinite past, the other in the infinite future; this is analogous to the fact that the conformal boundary of deSitter space itself consists of two 3-spheres in the infinite past and future.

Now, examining $\gamma_3$ more closely, we notice two crucial features. The first is initially surprising: this metric is {\em geodesically complete}, despite the fact that the naked singularities are eternal and so are ``still present at infinity". The reason for this is that while the spacetime has no event horizon, it does still have cosmological horizons which, as in any asymptotically deSitter spacetime, continuously ``contract" around a given point. Thus, at late times, the regions of future infinity which can be affected by the singularities becomes steadily smaller, until, at infinity itself, they have shrunk to two points. These two points do have to be excised, but the conformal freedom at infinity can be used to push the ``holes" off to an infinite distance. [One ``hole" is at $``t = \infty"$, the other at $``t = -\infty$".] In fact, this is an example of a classical construction due to Nomizu and Ozeki \cite{kn:nomizu}, whose theorem states that, given any Riemannian metric on any manifold, there exists a conformally related metric which is geodesically complete. The fact that this nakedly singular spacetime has a conformal infinity which is {\em complete} is a beautiful physical implementation of the Nomizu-Ozeki construction: the strong deSitter expansion has indeed pushed the holes off to an infinite spatial distance [at temporal infinity].

Now from the dS/CFT point of view, this is just as we would wish. For the correspondence, like its AdS/CFT counterpart, is supposed to relate a gravitational theory in the bulk to a {\em non-gravitational} theory on the boundary. A non-gravitational theory should be defined on a non-singular, geodesically complete space. The good behaviour of the geometry at infinity confirms our belief that the dS/CFT correspondence can be used here, in the sense that any misbehaviour of the boundary CFT cannot be ascribed to singularities of the boundary metric. [Related observations were made in \cite{kn:ghezelbash}.] For this reason, we shall henceforth only consider complete metrics on the boundary.

This discussion brings us to the second point: since, for reasons explained above, $t$ runs from $-\infty$ to $+\infty$, it is clear that the boundary is not compact. [Indeed, this is why the question of completeness is an issue here $-$ a compact manifold is geodesically complete with respect to any Riemannian [not Lorentzian] metric.] Before dealing with the consequences of this, let us contrast the situation here with pure deSitter space, with its metric
\begin{equation} \label{eq:dS}
g(dS_4) = -(1 - {{r^2}\over{L^2}}) {dt\otimes dt} + {(1 - {{r^2}\over{L^2}})}^{-1} {dr\otimes dr} + {r^2} {d{\Omega_2}}^2,
\end{equation}
where $d{\Omega_2}^2$ is the unit 2-sphere metric. Letting $r$ tend to infinity as above, we obtain 
\begin{equation} \label{eq:cylinder}
 dt \otimes dt + {L^2}{d{\Omega_2}}^2,
\end{equation}
which appears to be the metric on the non-compact cylinder $\bbr \times S^2$. Of course this is not correct: the future boundary is a three-sphere. The point is that the polar coordinates do not cover the origin or its antipode, and the excision of these two points from $S^3$ does indeed produce a topological cylinder; the deSitter expansion then implements a Nomizu-Ozeki completion, producing the metric \ref{eq:cylinder}. Evidently, the ``non-compactness" here is a mere coordinate effect. In the case of equation \ref{eq:gamma3}, however, the excisions are genuinely necessary, so that future infinity is genuinely non-compact; it is a three-dimensional manifold with topology $\bbr \times (\bbr^2/\Delta)$. Thus we see that while the {\em local} geometry of conformal infinity here is the same as that of the conformal infinity of a (Euclidean) hyperbolic AdS black hole spacetime, the {\em global} structure is quite different. But the methods we used to establish topological instability in the AdS case were based on the compactness of the boundary. The failure of asymptotic simplicity in this case produces, as the Andersson-Galloway theorem \cite{kn:galloway3} would lead us to expect, a non-compact boundary. It follows that {\em it is no longer clear that the scalar curvature on the boundary cannot be positive or zero everywhere}. We must therefore reconsider the consequences of sign conditions on the scalar curvature of the boundary, corresponding to singular asymptotically deSitter metrics like the one given by equation \ref{eq:dshyper} above.

The scalar curvature of the boundary metric used here [as representative of the conformal structure] is $-2/L^2$, so we can expect unstable behaviour for the CFT on the boundary. What is the bulk counterpart of this instability? A comparison with the AdS case suggests an answer: at extremely late times, beyond the horizon [in the uppermost ``diamond" of the Penrose diagram, where the metric is not static], we can expect spacelike branes to be ``emitted" [that is, to appear suddenly]. We interpret the boundary runaway as the dS/CFT dual of unstable production of these ``late branes", which are the dS/CFT analogue of the AdS/CFT ``large branes". Of course, the question now is this: can back-reaction bring the instability under control by modifying the geometry of the spatial sections, so that future conformal infinity actually has positive or zero scalar curvature, with late branes mediating the transition? As we saw, the methods used in section 2 to prove that the analogous AdS boundary cannot accept positive or zero scalar curvature do not work here, so the answer requires new techniques.

Again, these techniques do exist, but they are somewhat more abstruse than in the compact case, so we shall not attempt to summarize them; the reader may consult pages 313-326 of \cite{kn:lawson}. The main result we need may be stated as follows. Let $M$ be any compact enlargeable spin manifold, and let $g$ be any metric on $\bbr \times M$ such that the manifold is geodesically complete. [Recall that there is a physical motivation for requiring this last condition.] Then the scalar curvature of $g$ {\em cannot} be everywhere positive. Since ${\bbr^2}/\Delta$ is enlargeable [it has a metric of non-positive sectional curvature], we see that there is indeed no complete metric of positive scalar curvature on a three-dimensional  manifold with the topology that we have here, $\bbr \times (\bbr^2/\Delta)$, despite the non-compactness. In fact, we have a stronger result in this particular case. Suppose that the scalar curvature of a complete metric on this manifold satisfies $R \geq 0$ everywhere. Then a theorem of Kazdan \cite{kn:kazdan} states that the Ricci tensor, if it is not identically zero, can be used to deform the metric so that the scalar curvature becomes everywhere strictly positive. Thus $R \geq 0$ everywhere on $\bbr \times (\bbr^2/\Delta)$ implies that the Ricci tensor must vanish. But since $\bbr \times (\bbr^2/\Delta)$ is three-dimensional, this in turn means that the metric is flat. However, once again, this manifold has the wrong topology to be flat. We conclude that there is no complete metric with scalar curvature everywhere non-negative on this boundary manifold.

Once again, then, we are forced to conclude that no matter what effect late branes have on the spacetime geometry, their unstable production cannot be halted: it is simply impossible to deform a metric on $\bbr \times (\bbr^2/\Delta)$ in a way that would achieve this. Again, as in the previous section, processes which might change the topology of the spacelike sections would lead either to causality violation or to new singularities, and we would not normally interpret this as a sign that stability had been restored --- just the reverse. Clearly, such space-times  cannot occur either in dS/CFT or in whatever fundamental theory --- one hopes it is string or M theory --- underlies dS/CFT. If anything resembling dS/CFT is valid, then asymptotically deSitter S-brane spacetimes preserving R-symmetries are unstable.

\addtocounter{section}{4}
\section*{4.  Conclusion}

We have seen that Seiberg-Witten instability imposes a strong and unexpected constraint on the geometry of the event horizon of an asymptotically AdS black hole embedded in string theory. All {\em known} asymptotically AdS black holes are ``Seiberg-Witten unstable" in string theory, except those with horizons built up [by connected sums] from $S^1 \times S^2$ and quotients of spheres and tori. In fact, we believe that the word ``known" in this statement can be dropped; this depends on the purely topological conjectures that [a] all Eilenberg-MacLane spaces are enlargeable and [b] the Euclidean conformal boundary of a black hole spacetime necessarily inherits enlargeability from the horizon. Counterexamples to either of these conjectures, in the highly improbable event that any can be found, must have an extremely complex topology and are most unlikely to be of physical interest. Thus it seems that, for string theory on AdS backgrounds,
event horizon geometries cannot be much more complicated than in the asymptotically flat case. Surprisingly, we were able to establish this even if we allowed the large branes in the bulk to act back on the spacetime geometry. No matter how strong the back-reaction may be, the Seiberg-Witten criterion continues to hold if it holds initially. Only when conditions become so extreme that the {\em topology} of the spatial sections changes can there be any possibility of back-reaction bringing the large branes under control; but, by then, new singularities [or, worse still, causality violations] are generated, which presumably indicates that stability has been lost in any case.

We have also investigated the possibility of generalizing the asymptotically flat S-brane solutions of \cite{kn:gutperle1} to asymptotically deSitter S-brane spacetimes. Again we find, assuming that some kind of dS/CFT correspondence is valid, and assuming that the R-symmetries are preserved, that such solutions are unstable in a way which is immune to back-reaction. 

In \cite{kn:buchel1}\cite{kn:buchel2}, the consequences of including tachyonic matter in the effective dynamics of S-branes was studied, while maintaining the R-symmetries as usual. The main consequence is the replacement of the ``Milne horizon" by a spacelike singularity representing an S-brane. The spacetime otherwise evolves in an orderly way, though the timelike singularities are {\em not} resolved. As one would expect on the basis of the results obtained here, future conformal infinity remains non-spacelike. An asymptotically deSitter version of this spacetime would [because the timelike singularities are still eternal, and the hyperbolic part of the metric is still present] have the same kind of future conformal infinity as we discussed above, and we therefore predict that it would be unstable, with or without back-reaction. In fact, we predict that physically reasonable deSitter S-brane spacetimes can only be obtained by explicitly breaking the R-symmetries. Unfortunately, it is not known how to obtain explicit solutions in this case; the difficulties are discussed in \cite{kn:buchel1}\cite{kn:buchel2}. An alternative approach would be to try to generalize the S-brane solutions with flat transverse spaces, mentioned in \cite{kn:quevedo}. Unfortunately, the Seiberg-Witten criterion in the case where the boundary scalar curvature is zero is not fully understood. Perhaps a ``higher-derivative" approach [see for example \cite{kn:odintsov2} for references] can elucidate the higher-order terms in the large brane action.

\end{document}